\documentclass[12pt]{article}
\date{}
\usepackage{amsmath, amsfonts, amssymb, amsbsy, amsthm}
\usepackage{bbm, graphicx}
\usepackage{geometry}
\usepackage{parskip}
\usepackage{listings}
\usepackage{booktabs}
\usepackage{rotating}
\usepackage{natbib}
\usepackage{placeins} 
\usepackage{lscape}  
\usepackage{float}
\usepackage{multirow}
\usepackage{algpseudocode} 
\usepackage{adjustbox}
\usepackage{tcolorbox} 
\newtcolorbox{plainalgorithmbox}[1]{
    colback=white, 
    colframe=black, 
    fonttitle=\bfseries,
    title=#1,
    left=5pt, right=5pt, top=5pt, bottom=5pt,
    arc=0mm, 
    boxrule=0.8pt, 
}

\newcommand\IR{I\!\!R}

\newcommand{\toind}{\stackrel{d}{\to}}
\newcommand{\toinp}{\stackrel{p}{\to}}
\newcommand{\toas}{\stackrel{a.s.}{\to}}

\newcommand\argmin{\mathop{\rm argmin}}
\newcommand{\norm}[1]{\left\Vert#1\right\Vert}

\newcommand{\eps}{\varepsilon}

\newcommand{\ga}{\gamma}

\newcommand{\delg}{\nabla_{\gamma}}
\newcommand{\delgtr}{\nabla_{\gamma'}}
\newcommand{\delgg}{\nabla^2_{\gamma\gamma'}}

\newtheorem{thm}{Theorem}

\newtheorem{lem}{Lemma}

\title{Second-Order Least Squares Method for Statistical Inference in Linear Models with Asymmetric Errors}
\author{Sanha Kil, Yuliya V. Martsynyuk, Liqun Wang \\{\it Department of Statistics, 
University of Manitoba}}

\begin{document}

\maketitle

\begin{abstract}
In a linear regression model with normal random errors, the ordinary least squares estimator is the most efficient estimator. However, when the random errors have an asymmetric distribution, a more efficient estimator can be constructed using the higher moments of the data. In this paper, we use the second-order least squares method to obtain more accurate asymptotic confidence intervals and more powerful hypothesis tests in linear models. We also develop an efficient and stable numerical algorithm to calculate the proposed estimator. Our extensive simulation studies demonstrate the advantages of the proposed method. Two real data examples are presented.
\end{abstract}

\section{Introduction}

Asymmetrically distributed data arise in many fields, including environmental science, finance and economics, engineering, life-time and extreme data analysis, see e.g., \cite{begueria2006mapping, castro2022practical, green1984iteratively, wright1982quantitative, yousof2018new, nandy2025robust}. 
It is well-known that, in linear regression models, the ordinary least squares estimator (OLSE) is the best unbiased estimator and is most efficient if the model error has a normal distribution. However, if the model error has an asymmetric distribution, then the higher moments of the data can be used to construct more efficient estimators. 
One such example is the second-order least squares (SLS) estimator (SLSE) that uses the first two conditional moments of the response variable given the predictors. It was shown that this estimator is asymptotically more efficient than the OLSE if the model error has a non-zero third moment and that both estimators are asymptotically equivalent otherwise \citep{wang2008second}. Additionally, the SLSE attains the optimal semiparametric efficiency bound \citep{kim2012efficiency}. 
In this paper, we demonstrate, among other things, that the efficiency gain of the SLSE may lead to conclusions different from when the OLSE is used in real data analysis (Section 6).

While the asymptotic theory of the SLSE has been established for general nonlinear regression models (Wang and Leblanc, 2008), its more in-depth properties for statistical inferences in linear models have not been explored. In this paper, we fill in these gaps by investigating more properties of the SLSE in linear models and studying its performance in asymptotic confidence intervals and hypothesis testing problems. Moreover, the theoretical results of Wang and Leblanc (2008) were derived for general nonlinear models under some restrictive regularity conditions. Here we show that most of their conditions can be relaxed or removed in linear models, thus making the method more widely applicable. Another challenge with the SLSE is numerical computation, which has not been systematically explored in the literature. In Section 4, we develop an iteratively reweighted quasi-Newton method that is relatively easy to implement and efficient in practice, as well as numerically stable (Section 5.2). 

The SLS method has been used for parameter estimation problems, including in measurement error models \citep{wang2003estimation, wang2004estimation}, mixed effects models \citep{Wang2007unified}, censored linear models \citep{abarin2009second}, and dynamic models \citep{salamh2021a,  salamh2021b}. More recently, this method has also been used in optimal design problems \citep{gao2014new, bose2015optimal, yin2017optimal, yeh2021properties, he2022optimal, he2024optimal}.


\section{The SLS Estimator}

Consider the linear model $Y = X'\beta + \epsilon$, where $E(\epsilon|X)=0$ and $E(\epsilon^2\mid X)=\sigma^2$. Given the $i.i.d.$ random sample $(Y_i, X_i), i=1,2,...,n$, the OLSE $\hat\beta_{OLS}$ minimizes the ``first-order'' distance
\[
S_n(\beta) = \sum_{i=1}^n (Y_i-X_i'\beta)^2.
\]
The OLSE for $\sigma^2 $ is defined as
\[
\hat\sigma_{OLS}^2 = \frac{1}{n-p}S_n(\hat\beta_{OLS}),
\]
which is not obtained through an optimization process.

The SLSE is based on the first two (conditional) moments of $Y_i$ given $X_i$: $E(Y_i|X_i)= X_i'\beta_0$ and $E(Y_i^2|X_i) = (X_i'\beta_0)^2 + \sigma_0^2$, where $\beta_0$ and $\sigma_0^2$ are the true parameter values. Specifically, the SLSE for the vector of the unknown parameters $\gamma=(\beta',\sigma^2)'$ is defined as
\begin{equation}\label{gamma-hat}
\hat\gamma_{SLS} = \argmin Q_n(\gamma) := \sum_{i=1}^{n}\rho_i(\gamma)'W_i\rho_i(\gamma),
\end{equation}
where $\rho_i(\gamma) = (Y_i - X_i'\beta, Y_i^2 - (X_i'\beta)^2 - \sigma^2)'$ and $W_i=W(X_i)$ is a nonnegative definite weight matrix that may depend on $X_i$. 
\cite{wang2008second} showed that the most efficient SLSE is obtained by using the optimal weight matrix $W_i = E[\rho_i(\gamma_0)\rho_i'(\gamma_0) | X_i]^{-1}$, with $\gamma_0=(\beta_0',\sigma_0^2)'$, which has the following form for linear models:
\begin{equation}
W_i = \eta \begin{pmatrix}
\mu_4 - \sigma_0^4  + 4\mu_3(X_i'\beta_0) + 4\sigma_0^2(X_i'\beta_0)^2
&-\mu_3 - 2\sigma_0^2(X_i'\beta_0)\\
-\mu_3 - 2\sigma_0^2(X_i'\beta_0)
&\sigma_0^2
\end{pmatrix},
\end{equation}
where $\mu_j=E(\eps^j|X)$, $j=3,4$ and $\eta = [\sigma_0^2(\mu_4-\sigma_0^4)-\mu_3^2]^{-1}$. In this paper, we will use the term SLSE to refer to the most efficient version of the estimator. 

Further, \cite{wang2008second} established  consistency and asymptotic normality of the SLSE in nonlinear models under general conditions. Their conditions can be substantially simplified for linear models as follows.

\begin{description}
    \item[A1] The error $\eps$ has finite constant moments $\mu_j=E(\eps^j|X)$, $j=3,4,$ and $\sigma_0^2(\mu_4-\sigma_0^4)-\mu_3^2 \neq 0$.

    \item[A2] The predictor $X$ has a finite fourth moment $E\norm{X}^4 < \infty$.

    \item[A3] The  true parameter value $\gamma_0=(\beta_0',\sigma_0^2)'\in int(\Gamma)$ and the parameter space $\Gamma \subset \mathbb{R}^{p+1}$ is compact.
\end{description}

Let  $\mu_x = E(X_i)$ and $M_{xx} = E(X_iX_i')$. Then we have the following results, the proofs of which are given in Appendix.

\begin{thm}\label{theorem1}
Under assumptions A1-A3, as $n\to\infty$, the SLSE satisfies
\begin{enumerate}
\item[1)]
$\hat\gamma_{SLS}\toas\gamma$;
\item[2)]
$\sqrt{n}(\hat\gamma_{SLS} - \gamma) \toind N(0, C)$, where
\begin{equation}\label{AV-gamma}
    C = \begin{pmatrix}
    V(\hat\beta_{SLS}) & \xi V(\hat\sigma^2_{SLS})M_{xx}^{-1}\mu_x \\
    \xi V(\hat\sigma^2_{SLS})\mu_x'M_{xx}^{-1} &V(\hat\sigma^2_{SLS}) 
    \end{pmatrix},~~\mbox{with}~~\xi = \mu_3^2/[\sigma_0^2(\mu_4-\sigma_0^4)],
\end{equation}
\begin{equation}\label{AV-betta}
V(\hat\beta_{SLS}) = \sigma_0^2(1-\xi)(M_{xx} - \xi\mu_x\mu_x')^{-1},
\end{equation}
and
\begin{equation}\label{AV-betta}
V(\hat\sigma^2_{SLS}) = \frac{(\mu_4-\sigma_0^4)(1-\xi)}{1 - \xi \mu_x' M_{xx}^{-1} \mu_x}.
\end{equation}
\end{enumerate}
\end{thm}

From Theorem \ref{theorem1} we see that when $\mu_3=0$ and hence $\xi=0$, the SLSEs $\hat\beta_{SLS}$ and $\hat\sigma^2_{SLS}$ have the same asymptotic variances as the respective OLSEs: $V(\hat\beta_{SLS})=V(\hat\beta_{OLS}) = \sigma_0^2M_{xx}^{-1}$ and $V(\hat\sigma^2_{SLS})=V(\hat\sigma^2_{OLS}) = \mu_4 - \sigma_0^4$. However, the following theorem shows that in general the SLSE is asymptotically more efficient than the OLSE when $\mu_3\neq 0$.

\begin{thm} Suppose $\mu_3\neq 0$. Then, under the conditions of Theorem 1,
\begin{enumerate}
\item[1)]
The matrix
$V(\hat\beta_{OLS}) - V(\hat\beta_{SLS})$ is positive definite if and only if $\mu_x'M^{-1}_{xx}\mu_x \neq 1$; and is nonnegative definite with rank $\dim(\beta) -1$ if $\mu_x'M^{-1}_{xx}\mu_x = 1$.
\item[2)]
$V(\hat\sigma^2_{OLS}) - V(\hat\sigma^2_{SLS})\geq 0$ with equality holding if and only if $\mu_x'M^{-1}_{xx}\mu_x = 1$.
\end{enumerate}
\end{thm}

From Theorem 2 we can see that the efficiency gain of the SLSE depends on the value of $ \mu_x'M^{-1}_{xx}\mu_x\in [0,1]$ (see Appendix C). For example, the maximum gain is achieved when the data on $X$ are centerred, so that $\mu_x=0$, in which case $V(\hat\beta_{SLS}) = (1-\xi) V(\hat\beta_{OLS})$ and $V(\hat\sigma^2_{SLS}) = (1-\xi)V(\hat\sigma^2_{OLS})$.

On the other hand, if the model has an intercept term, i.e., for the first predictor variable $P(X_1=1)=1$, then $M_{xx}e_1 = \mu_x$ and therefore $\mu_x'M^{-1}_{xx}\mu_x = 1$, where $e_1$ is the unit vector with the first element one and others zero. In this case, the matrix $V(\hat\beta_{OLS}) - V(\hat\beta_{SLS})$ have all positive but one zero eigenvalues. In particular, using the block matrix inversion formula we obtain an alternative expression
\begin{equation}\label{AV-beta-2}
V(\hat\beta_{SLS}) = \sigma_0^2\bigl[(1-\xi)M_{xx}^{-1} + \xi e_1e_1'\bigr].
\end{equation}
However, in the case where the model does not have intercept, then $\mu_x'M^{-1}_{xx}\mu_x <1$ and, therefore, all elements of $\hat\beta_{SLS}$ are asymptotically more efficient than the corresponding those of $\hat\beta_{OLS}$.

\section{Statistical Inference}

Although the SLS method has been used in point estimation and experimental design problems in the literature, it has not been used in confidence intervals and hypothesis testing problems. In this section, we use the asymptotic theory of the previous section to construct Wald statistics for statistical inferences. Specifically, by Theorem 1, for each $j=1,2,...,p$,  the $j$-th element $\hat{\beta}_{SLS,j}$ of $\hat\beta_{SLS}$ satisfies $\hat s_{j}^{-1}(\hat{\beta}_{SLS,j} - \beta_j)\toind N(0, 1)$, where $\hat s_{_j}^2$ is the $j$-th diagonal element of the estimated version of the matrix $V(\hat\beta_{SLS})$ of (4), where all moments and true parameter values are replaced with the sample moments and SLSEs, respectively.

Therefore, for any $0<\alpha<1$, a level $1-\alpha$ asymptotic confidence interval for $\beta_j$ is
\begin{equation}
    \left[ \hat{\beta}_{SLS,j} - z_{\alpha/2} \hat s_{j},\ 
       \hat{\beta}_{SLS,j} + z_{\alpha/2} \hat s_{j} \right],
\end{equation}
where $z_{\alpha/2}$ is the upper $\alpha/2$ quantile of the standard normal distribution.

Similarly, for testing the  hypotheses 
\begin{equation}
H_0 : \beta_j = \beta_{j0}\quad\mbox{vs.}\quad H_1 : \beta_j \ne \beta_{j0},
\end{equation}
where $\beta_{j0}$ is known, the Wald test statistic is 
\[
Z_j = \frac{\hat{\beta}_{SLS,j} - \beta_{j0}}{\hat s_{j}}.
\]
Hence, for a given significance level $0<\alpha<1$, $H_0$ will be rejected if $|Z_j| \ge z_{\alpha/2}$.

\section{Reweighted Quasi-Newton Method}

Although the SLSE has desirable statistical properties under skewed error distributions, its numerical computation is not straightforward, as it does not have a closed-form formula as the OLSE.
In this section, we develop a numerical algorithm to calculate the SLSE in general linear models, based on the Newton method for minimization of the objective function $Q_n(\gamma)$ in $(\ref{gamma-hat})$.

First, the standard Newton procedure is based on the iterative equation
\begin{equation}
\gamma^{(t+1)} = \gamma^{(t)} - H(\gamma^{(t)})^{-1}G(\gamma^{(t)}),\quad t = 0,1,2,... 
\end{equation}
where $G(\gamma)$ and $H(\gamma)$ are the gradient and Hessian matrix of $Q_n(\gamma)$, respectively:
\begin{align}
G(\gamma) &= \delg Q_n(\gamma)
= 2\sum_{i=1}^{n}\delg\rho_i'(\gamma) W_i \rho_i(\gamma)
\end{align} 
and
\begin{align}
H(\gamma) &= \delgg Q_n(\gamma)
= 2\sum_{i=1}^{n} \delg\rho_i'(\gamma) W_i \delgtr\rho_i(\gamma) + (\rho_i'(\ga)W_i\otimes I_{p+1})\delgtr\mbox{vec}(\delg\rho_i'(\ga)).
\end{align} 

However, the weight matrix $W_i$ of (2) in the gradient and Hessian depends on the unknown parameters $\gamma$, $\mu_3$ and $\mu_4$. To overcome this complication, we propose a numerically simpler and more efficient quasi-Newton method, where these parameters are updated in each iteration by using the current value $\hat\gamma^{(t)}$ and the corresponding residuals $e_i^{(t)} = y_i-x_i'\hat\beta^{(t)}$. 
This so-called reweighted Newton update equation is given by (9) with 
\begin{equation}\label{gradient}
G(\gamma) = -\eta\sum_{i=1}^{n}
\begin{pmatrix}
[(\mu_4 - \sigma^4)(y_i - x_i'\beta) -\mu_3(y_i - x_i'\beta)^2 +  \mu_3\sigma^2 ]x_i\\
\sigma^2(y_i - x_i'\beta)^2 - \sigma^4 - \mu_3(y_i - x_i'\beta) 
\end{pmatrix},
\end{equation} 
\begin{equation}\label{hessian}
H(\gamma) = \eta \sum_{i=1}^{n} 
\begin{pmatrix}
[\mu_4 + \sigma^4 - 2\sigma^2(y_i - x_i'\beta)^2 + 2\mu_3 (y_i - x_i'\beta)]x_ix_i'    &-\mu_3 x_i\\
-\mu_3 x_i'   &\sigma^2
\end{pmatrix},
\end{equation} 
and $\mu_j = n^{-1}\sum_{i=1}^n \bigl(y_i-x_i'\beta^{(t)}\bigr)^j$, $j=3,4$. The derivation of (\ref{gradient}) and (\ref{hessian}) are given in Appendix D.

Another challenge is that in the case of small samples, the so-called ``bad samples'' may cause the Hessian matrix to be near singular.
To address this issue, we implemented adaptive ridge regularization and Armijo backtracking line search, which assure the stability of the Hessian and prevent divergence of the algorithm.

First, we initialize $\lambda = 10^{-8}$ and increase it over iteration until $\tilde{H}(\gamma^{(t)}) = H(\gamma^{(t)})+\lambda I_p$ becomes positive definite. This way, the descent direction is computed as $d^{(t)}=-\tilde{H}(\gamma^{(t)})G(\gamma^{(t)})$.
Further, to introduce stronger stability and to increase convergence speed, Armijo backtracking line search is utilized as
\[
\gamma^{(t+1)} = \gamma^{(t)} + \alpha_k d^{(t)},
\]
where $\alpha_{k} \in (0,1]$ is the step size. We start with $\alpha_0 = 1$ and decrease it until two conditions (Armijo Conditions) are met:
\begin{enumerate}
\item
Boundary constraint: $\gamma_p^{(t)} >0$;
\item
Sufficient decrease: 
$Q_n(\gamma^{(t)}+\alpha_kd^{(t)}) \leq Q_n(\gamma^{(t)}) + c\alpha_k G(\gamma^{(t)})^{'}d^{(t)}$,
where c is a small constant ensuring a sufficient reduction.
\end{enumerate}

Finally, we use the OLSE as the starting point of the iteration process.
The algorithm of this reweighted quasi-Newton procedure is given below. The numerical stability of the algorithm will be illustrated for two simulation models in Section 5.2.

\begin{plainalgorithmbox}{Algorithm : Reweighted Quasi-Newton Method with Line Search}
\begin{algorithmic}[1]
    \State \textbf{Input:} Data $(Y, X)$, Initial estimate $\gamma^{(0)}$, Convergence tolerance $\delta$, Max iterations $K_{max}$.
    \State \textbf{Initialize:} $t \leftarrow 0$, $\gamma^{(t)} \leftarrow \gamma^{(0)}$.

    \While{$||\gamma^{(t)} - \gamma^{(t-1)}|| > \delta$ \textbf{and} $t < K_{max}$}
        \State Compute Gradient $G(\gamma^{(t)})$ and Hessian $H(\gamma^{(t)})$ using estimated moments $\hat{\mu}_j$.

        \State \Comment{\textbf{Adaptive Ridge Regularization}}
        \State Set $\lambda \leftarrow 10^{-8}$.
        \State \textbf{Repeat}
            \State $\tilde{H} \leftarrow H^{(t)} + \lambda I_p$.
            \State \textbf{If} $\tilde{H}$ is Positive Definite (via Cholesky): \textbf{Break}
            \State $\lambda \leftarrow 10 \times \lambda$.
        \State Compute Descent Direction $d^{(t)} \leftarrow -\tilde{H}^{-1} G(\gamma^{(t)})$.

        \State \Comment{\textbf{Armijo Backtracking Line Search}}
        \State Set $\alpha \leftarrow 1$.
        \State \textbf{Repeat}
            \State Candidate $\gamma_{new} \leftarrow \gamma^{(t)} + \alpha d^{(t)}$.
            \State \textbf{If} $\gamma_{new}[p] \le 0$ (\textit{variance constraint violated}): $\alpha \leftarrow 0.5 \alpha$; \textbf{Continue}
            \State \textbf{If} $Q_n(\gamma_{new}) \le Q_n(\gamma^{(t)}) + c_1 \alpha G(\gamma^{(t)})^{'} d^{(t)}$: \textbf{Break} (\textit{Sufficient Decrease met})
            \State $\alpha \leftarrow 0.5 \times \alpha$.
        \State \textbf{Until} $\alpha$ reaches minimum tolerance.

        \State \Comment{\textbf{Update and Iterate}}
        \State $\gamma^{(t+1)} \leftarrow \gamma_{new}$.
        \State $t \leftarrow t + 1$.
    \EndWhile

    \State \textbf{Output:} $\hat{\gamma} \leftarrow \gamma^{(t)}$.
\end{algorithmic}
\end{plainalgorithmbox}

\section{Simulation Studies}

In this section, we investigate numerically the finite-sample performance of the SLSE $\hat\beta_{SLS,j}$ vs.\ the OLSE $\hat\beta_{OLS,j}$ for each of the components $\beta_j$ of $\beta$, the asymptotic CIs for $\beta_j$ based on the asymptotic normality of $\hat\beta_{SLS,j}$ from Theorem 1 vs.\ the well-known CIs based on $\hat\beta_{OLS,j}$, as well as the respective asymptotic tests of the hypotheses in (8) using the corresponding Wald test statistics. In view of that $\hat\sigma^2_{SLS}$ converges to $\sigma^2$ much slower compared to the convergence of $\hat\beta_{SLS}$ to $\beta$, comparison of the $\hat\sigma^2_{SLS}$ vs.\ $\hat\sigma^2_{OLS}$ based inference results is omitted in this study.

\subsection{Simulation models}
For our simulation studies, we consider the following two models:

\begin{description}
    \item[Model 1:] $Y = \beta_0X_1 + \beta_1 X_2 + \beta_2 X_3 + \eps$ with $\beta = (0.5, 1, -1.5)'$, where $X_1 =1$, $X_2\sim Unif(1,20)$, $X_3\sim\mathcal{N}(5, 4)$, and $X\perp\!\!\!\perp\epsilon$.
    \item[Model 2:] $Y = \beta_0X_1 + \beta_1 X_2 + \beta_2 X_3 + \eps$ with $\beta = (0.5, 1, -1.5)'$, where $X_1 \sim Beta(2,5)$, $(X_2,X_3)'\sim\mathcal{N}\left((2,0)', [1, 0.5; 0.5, 1]\right)$, and $X\perp\!\!\!\perp\epsilon$.
\end{description}

For each of the models, we consider the following three error distributions with the means of zero and the variances of one: $\eps\sim N(0, 1)$; $\eps\sim Exp(1) - 1$, called std-Exp$(1)$ hereafter, with the skewness of 2; and $\eps\sim ({\rm Weibull}(0.6 , 1) - \Gamma(1 + 1/0.6)) /  \sqrt{(\Gamma(1 + 2/0.6) - (\Gamma(1 + 1/0.6))^2)}$, abbreviated as std-Weibull$(0.6,1)$, with the skewness of 4.59.

In our simulation studies, we use the sample sizes $n = 100$, 300, 500, 1000 and, for some tables, $n=50$, 5000, 10000, 20000, 50000, and $m = 5000$ Monte Carlo replications.

We use the large-sample standard errors (ASE) and empirical standard errors (SE) to compare $\hat\beta_{SLS,j}$ and $\hat\beta_{OLS,j}$ for each component $\beta_j$ of $\beta$. The performance of the respective CIs is studied based on their empirical finite-sample coverage probabilities (CP) and the ratio of their empirical expected lengths, namely,
\begin{equation}
R_n = \frac{\text{average length of $m$ $\hat\beta_{OLS,j}$ based CIs for a sample size $n$}}{\text{average length of $m$ $\hat\beta_{SLS,j}$ based CIs for a sample size $n$}}.
\end{equation}
The performance of the $\hat\beta_{OLS,j}$ and $\hat\beta_{SLS,j}$ based Wald tests for the hypotheses in (8) is evaluated through their empirical power functions.

\subsection{Numerical stability of the SLSE}

Before studying the performance of the SLSE for $\beta$ in statistical inference, we address numerical stability of our proposed reweighted quasi-Newton algorithm for computing $\hat\beta_{SLS}$.

Under the standard reweighted Newton–Raphson algorithm (without regularization or line search), a large proportion of simulation runs resulted in failure due to instability in Hessian inversion. This issue was particularly severe in Model 2 for smaller sample sizes, where the rate of such “bad samples” reached as high as 37.14\% under the std-$\mathrm{Weibull}(0.6,1)$ error distribution for $n=50$ and was at 5.96\% for this skewed distribution even for $n=1000$, as shown in Table 1.

By introducing the reweighted quasi-Newton method with an Armijo-type line search, these failures were effectively eliminated. As seen from Table 1, across all simulation settings for both Models 1 and 2, the bad sample rate was reduced dramatically, dropping to essentially 0\%. This highlights the importance of regularization and adaptive step-size control in ensuring reliable implementation of the SLSE, particularly in smaller samples and under asymmetric error distributions.

\begin{table}[htbp]
\centering
\caption{Bad sample rates when computing the SLSE using the standard reweighted Newton method and Section 4 reweighted quasi-Newton method.}
\medskip
\label{tab:bad-sample-rate}
\small
\setlength{\tabcolsep}{3pt}
\renewcommand{\arraystretch}{1.12}
\resizebox{\textwidth}{!}{%
\begin{tabular}{@{}llccccc ccccc@{}}
\toprule
& & \multicolumn{5}{c}{Reweighted Newton}
& \multicolumn{5}{c}{Reweighted quasi--Newton} \\
\cmidrule(lr){3-7} \cmidrule(lr){8-12}
Model & Error distribution
& $n=50$ & $n=100$ & $n=300$ & $n=500$ & $n=1000$
& $n=50$ & $n=100$ & $n=300$ & $n=500$ & $n=1000$ \\
\midrule
Model 1 & N(0,1)
& 13.76\% & 3.47\% & $<0.01\%$ & 0\% & 0\%
& 0\% & 0\% & 0\% & 0\% & 0\% \\

Model 1 & std-Exp(1)
& 8.22\% & 0.73\% & $<0.01\%$ & 0\% & 0\%
& 0\% & 0\% & 0\% & 0\% & 0\% \\

Model 1 & std-Weibull$(0.6,1)$
& 2.81\% & $<0.01\%$ & $<0.01\%$ & 0\% & 0\%
& $<0.01\%$ & 0\% & 0\% & 0\% & 0\% \\

\midrule
Model 2 & N(0,1)
& 16.54\% & 7.15\% & 0.02\% & 0\% & 0\%
& 0\% & 0\% & 0\% & 0\% & 0\% \\

Model 2 & std-Exp(1)
& 28.41\% & 13.22\% & 2.64\% & 0.52\% & 0.02\%
& 0\% & 0\% & 0\% & 0\% & 0\% \\

Model 2 & std-Weibull$(0.6,1)$
& 37.14\% & 30.37\% & 20.43\% & 14.43\% & 5.96\%
& $<0.01\%$ & $<0.01\%$ & 0\% & 0\% & 0\% \\
\bottomrule
\end{tabular}%
}
\end{table}

\subsection{Comparison of the OLS and SLS based statistical inference}

Tables 2 and 3 below for Models 1 and 2, respectively, with three different error distributions, contain the values of the OLSE and SLSE, their respective ASEs and SEs, the CPs of the corresponding 95\% CIs, and the ratio $R_n$ of (14), for each $\beta_j$. 

Notably, across all simulation settings for both models, $R_n$ is greater than 1, indicating that the SLSE based CI is shorter on average than the OLSE based CI. For the N$(0,1)$ distributed errors, $R_n$ is very close to one, which agrees with the fact that the OLSE and SLSE estimators for $\beta_j$ are equally efficient for symmetrically distributed errors, as noted right after Theorem 1. However, for the std-Exp$(1)$ and more skewed std-Weibull$(0.6,1)$ errors, the SLSE based CI is significantly shorter than the OLSE CI, more so for the std-Weibull$(0.6,1)$ errors. Moreover, as illustrated in Table 4, due to the very slow rate of convergence of $R_n$ for these two error distributions, this advantage of the SLSE based CI continues to hold at least up to the sample size of 50000. Furthermore, we can also see that the efficiency gain of the SLSE is greater in Model 2 without the intercept, which is consistent with our earlier discussions after Theorem 2 in Section 2.

As for the empirical coverages of the OLSE and SLSE based CIs, they appear to converge fast for both models. Moreover, 
they are essentially on par and are both close enough to 95\% for $n\geq 300$ in Tables 2 and 3.
For $n=100$, the coverage of the SLSE based CI is somewhat lower than that of the OLSE based CI, by at most 1.12\% in Table 2 and 2.4\% in Table 3, while not getting less than 92.94\% for Model 1 and 92.28\% for Model 2.

The superior asymptotic efficiency of the SLSE for skewed error distributions can also be seen through its lower ASEs for the std-Exp(1) and std-Weibull(0.6,1) errors in Tables 2 and 3. The empirical SEs of the SLSE, which align closely with its ASEs, are also lower than the SEs of the OLSE for these error distributions, which indicates a higher precision of the SLSE in finite samples.

Tables 5 and 6 contain the empirical power function comparison for the OLSE and SLSE based tests of the hypotheses in (8) for Models 1 and 2, respectively. Both tests have well-controlled empirical type I errors that are close enough to the nominal 5\% for $n\geq 300$, except for a few cases for the std-Weibull$(0.6,1)$ errors and $n=300$ and 500 for Model 2, where one or both errors are between 6.2\% and 6.8\%. For $n=100$ and all three parameters $\beta_0$, $\beta_1$, and $\beta_2$ for both models, the type I errors of the SLSE based test are somewhat higher than those of the OLSE based test, by at most 1.6\% in Table 5 and 2.04\% in Table 6, where the type I errors of the SLSE based test do not exceed 6.9\% and 7.86\%, respectively. 

In order to compare the powers of the OLSE and SLSE based tests under the two-sided alternative hypotheses in (8), we consider two values of $\beta_{j0}$ for each of $\beta_j$. Notably, the SLSE based test is significantly more powerful than the OLSE based one across all settings in Tables 5 and 6. Both test powers increase with sample size and are higher for the values of $\beta_{j0}$ that are further away from the respective true parameter values. For $n=1000$ for Model 1, both powers are essentially at 100\% for $\beta_{j0}$ that are further away from the true $\beta_j$, while for the other value of $\beta_{j0}$, Table 5 shows that the power of the OLSE based test is still behind that of the SLSE based one, by as much as 30\% for  the std-Weibull(0.6,1) errors and $\beta_{00}=0.72$. Compared to Model 1, both the OLSE and SLSE based tests are less powerful across all settings for Model 2, where the power convergence for both tests appears to be slower and only the SLSE based test power reaches almost 100\% at $n=1000$ and only for $\beta_{00}=0.94$ and $\beta_{20}=-1.40$. Just like for Model 1, the power superiority of the SLSE based test is still considerable across all sample sizes, where the latter test is as much as 34.6\%  more powerful for the std-Weibull$(0.6,1)$ errors and $\beta_{20}=-1.45$.

\clearpage
\begin{table}[!p]
\centering
\caption{OLSE vs.\ SLSE, and 95\% OLS CI vs.\ 95\% SLS CI for Model 1. }
\label{tab:model1-main}
\scriptsize
\setlength{\tabcolsep}{2.0pt}
\renewcommand{\arraystretch}{0.72}
\resizebox{\textwidth}{!}{%
\begin{tabular}{@{}ll l cccc cccc c@{}}
\toprule
Error distribution & $n$ & Parameter
& \multicolumn{4}{c}{OLS}
& \multicolumn{4}{c}{SLS}
& $R_n$ \\
\cmidrule(lr){4-7} \cmidrule(lr){8-11}
& & & OLSE & ASE & SE & CP & SLSE & ASE & SE & CP &  \\
\midrule

\multirow{12}{*}{\textbf{N(0,1)}}
& \multirow{3}{*}{$100$} & $\beta_0$ & 0.5041 & 0.3238 & 0.3236 & 0.9432 & 0.5044 & 0.3147 & 0.3287 & 0.9320 & 1.0292 \\
& & $\beta_1$ & 1.0001 & 0.0175 & 0.0178 & 0.9400 & 1.0001 & 0.0170 & 0.0182 & 0.9294 & 1.0307 \\
& & $\beta_2$ & -1.5006 & 0.0508 & 0.0505 & 0.9430 & -1.5007 & 0.0493 & 0.0513 & 0.9318 & 1.0307 \\
\cmidrule(lr){2-12}
& \multirow{3}{*}{$300$} & $\beta_0$ & 0.4999 & 0.1854 & 0.1852 & 0.9481 & 0.4998 & 0.1836 & 0.1857 & 0.9448 & 1.0095 \\
& & $\beta_1$ & 0.9999 & 0.0100 & 0.0101 & 0.9467 & 0.9999 & 0.0099 & 0.0101 & 0.9430 & 1.0100 \\
& & $\beta_2$ & -1.4999 & 0.0290 & 0.0290 & 0.9515 & -1.4999 & 0.0287 & 0.0291 & 0.9471 & 1.0100 \\
\cmidrule(lr){2-12}
& \multirow{3}{*}{$500$} & $\beta_0$ & 0.5005 & 0.1435 & 0.1456 & 0.9457 & 0.5005 & 0.1427 & 0.1460 & 0.9435 & 1.0057 \\
& & $\beta_1$ & 1.0000 & 0.0078 & 0.0077 & 0.9505 & 1.0000 & 0.0077 & 0.0077 & 0.9477 & 1.0060 \\
& & $\beta_2$ & -1.4999 & 0.0224 & 0.0228 & 0.9447 & -1.4999 & 0.0223 & 0.0229 & 0.9428 & 1.0060 \\
\cmidrule(lr){2-12}
& \multirow{3}{*}{$1000$} & $\beta_0$ & 0.5014 & 0.1014 & 0.1019 & 0.9503 & 0.5014 & 0.1011 & 0.1020 & 0.9499 & 1.0028 \\
& & $\beta_1$ & 0.9999 & 0.0055 & 0.0055 & 0.9529 & 0.9999 & 0.0055 & 0.0055 & 0.9519 & 1.0030 \\
& & $\beta_2$ & -1.5001 & 0.0158 & 0.0159 & 0.9488 & -1.5001 & 0.0158 & 0.0159 & 0.9474 & 1.0030 \\

\midrule
\multirow{12}{*}{\textbf{std-Exp(1)}}
& \multirow{3}{*}{$100$} & $\beta_0$ & 0.4957 & 0.3216 & 0.3278 & 0.9483 & 0.4972 & 0.2245 & 0.2282 & 0.9448 & 1.4329 \\
& & $\beta_1$ & 1.0000 & 0.0174 & 0.0175 & 0.9497 & 0.9999 & 0.0115 & 0.0117 & 0.9480 & 1.5165 \\
& & $\beta_2$ & -1.4994 & 0.0504 & 0.0518 & 0.9432 & -1.4995 & 0.0332 & 0.0342 & 0.9430 & 1.5165 \\
\cmidrule(lr){2-12}
& \multirow{3}{*}{$300$} & $\beta_0$ & 0.5012 & 0.1851 & 0.1858 & 0.9467 & 0.4994 & 0.1327 & 0.1336 & 0.9495 & 1.3945 \\
& & $\beta_1$ & 0.9999 & 0.0100 & 0.0100 & 0.9486 & 1.0000 & 0.0068 & 0.0069 & 0.9489 & 1.4702 \\
& & $\beta_2$ & -1.5000 & 0.0290 & 0.0292 & 0.9466 & -1.4999 & 0.0197 & 0.0199 & 0.9504 & 1.4701 \\
\cmidrule(lr){2-12}
& \multirow{3}{*}{$500$} & $\beta_0$ & 0.4976 & 0.1432 & 0.1450 & 0.9473 & 0.4992 & 0.1036 & 0.1038 & 0.9484 & 1.3822 \\
& & $\beta_1$ & 1.0001 & 0.0077 & 0.0078 & 0.9487 & 1.0000 & 0.0053 & 0.0054 & 0.9475 & 1.4548 \\
& & $\beta_2$ & -1.4998 & 0.0224 & 0.0225 & 0.9479 & -1.5000 & 0.0154 & 0.0154 & 0.9502 & 1.4548 \\
\cmidrule(lr){2-12}
& \multirow{3}{*}{$1000$} & $\beta_0$ & 0.4992 & 0.1013 & 0.1004 & 0.9501 & 0.4995 & 0.0739 & 0.0742 & 0.9508 & 1.3710 \\
& & $\beta_1$ & 1.0000 & 0.0055 & 0.0055 & 0.9488 & 1.0000 & 0.0038 & 0.0038 & 0.9476 & 1.4408 \\
& & $\beta_2$ & -1.4999 & 0.0158 & 0.0158 & 0.9473 & -1.5000 & 0.0110 & 0.0110 & 0.9512 & 1.4408 \\

\midrule
\multirow{12}{*}{\textbf{std-Weibull(0.6,1)}}
& \multirow{3}{*}{$100$} & $\beta_0$ & 0.5014 & 0.3130 & 0.3245 & 0.9435 & 0.4994 & 0.1801 & 0.1869 & 0.9357 & 1.7374 \\
& & $\beta_1$ & 0.9998 & 0.0169 & 0.0177 & 0.9454 & 1.0000 & 0.0086 & 0.0089 & 0.9499 & 1.9546 \\
& & $\beta_2$ & -1.4999 & 0.0490 & 0.0510 & 0.9478 & -1.4999 & 0.0251 & 0.0264 & 0.9425 & 1.9548 \\
\cmidrule(lr){2-12}
& \multirow{3}{*}{$300$} & $\beta_0$ & 0.4965 & 0.1827 & 0.1871 & 0.9448 & 0.4979 & 0.1126 & 0.1149 & 0.9429 & 1.6217 \\
& & $\beta_1$ & 1.0002 & 0.0099 & 0.0099 & 0.9505 & 1.0001 & 0.0055 & 0.0055 & 0.9548 & 1.7843 \\
& & $\beta_2$ & -1.4999 & 0.0285 & 0.0292 & 0.9478 & -1.5000 & 0.0160 & 0.0163 & 0.9496 & 1.7843 \\
\cmidrule(lr){2-12}
& \multirow{3}{*}{$500$} & $\beta_0$ & 0.5010 & 0.1424 & 0.1443 & 0.9478 & 0.5011 & 0.0904 & 0.0906 & 0.9534 & 1.5752 \\
& & $\beta_1$ & 1.0000 & 0.0077 & 0.0078 & 0.9464 & 1.0000 & 0.0045 & 0.0046 & 0.9493 & 1.7178 \\
& & $\beta_2$ & -1.5000 & 0.0223 & 0.0225 & 0.9500 & -1.5001 & 0.0130 & 0.0130 & 0.9521 & 1.7178 \\
\cmidrule(lr){2-12}
& \multirow{3}{*}{$1000$} & $\beta_0$ & 0.5000 & 0.1010 & 0.1010 & 0.9469 & 0.5008 & 0.0662 & 0.0666 & 0.9489 & 1.5269 \\
& & $\beta_1$ & 1.0000 & 0.0055 & 0.0055 & 0.9515 & 1.0000 & 0.0033 & 0.0033 & 0.9534 & 1.6501 \\
& & $\beta_2$ & -1.5000 & 0.0158 & 0.0159 & 0.9495 & -1.5001 & 0.0096 & 0.0097 & 0.9483 & 1.6502 \\

\bottomrule
\end{tabular}%
}
\end{table}

\begin{table}[!p]
\centering
\caption{OLSE vs.\ SLSE, and 95\% OLS CI vs.\ 95\% SLS CI for Model 2.}
\label{tab:model2-main}
\scriptsize
\setlength{\tabcolsep}{2.0pt}
\renewcommand{\arraystretch}{0.72}
\resizebox{\textwidth}{!}{%
\begin{tabular}{@{}ll l cccc cccc c@{}}
\toprule
Error distribution & $n$ & Parameter
& \multicolumn{4}{c}{OLS}
& \multicolumn{4}{c}{SLS}
& $R_n$ \\
\cmidrule(lr){4-7} \cmidrule(lr){8-11}
& & & OLSE & ASE & SE & CP & SLSE & ASE & SE & CP &  \\
\midrule

\multirow{12}{*}{\textbf{N(0,1)}}
& \multirow{3}{*}{$100$} & $\beta_0$ & 0.4916 & 0.5185 & 0.5271 & 0.9442 & 0.4902 & 0.5023 & 0.5329 & 0.9328 & 1.0321 \\
& & $\beta_1$ & 1.0008 & 0.0778 & 0.0785 & 0.9464 & 1.0010 & 0.0755 & 0.0794 & 0.9356 & 1.0307 \\
& & $\beta_2$ & -1.4972 & 0.1088 & 0.1107 & 0.9402 & -1.4974 & 0.1053 & 0.1120 & 0.9304 & 1.0328 \\
\cmidrule(lr){2-12}
& \multirow{3}{*}{$300$} & $\beta_0$ & 0.4996 & 0.2961 & 0.2960 & 0.9462 & 0.4991 & 0.2928 & 0.2980 & 0.9424 & 1.0112 \\
& & $\beta_1$ & 0.9999 & 0.0444 & 0.0444 & 0.9489 & 1.0000 & 0.0439 & 0.0446 & 0.9445 & 1.0107 \\
& & $\beta_2$ & -1.4999 & 0.0621 & 0.0628 & 0.9477 & -1.4999 & 0.0614 & 0.0631 & 0.9444 & 1.0114 \\
\cmidrule(lr){2-12}
& \multirow{3}{*}{$500$} & $\beta_0$ & 0.4992 & 0.2290 & 0.2300 & 0.9461 & 0.4989 & 0.2275 & 0.2309 & 0.9449 & 1.0068 \\
& & $\beta_1$ & 1.0002 & 0.0344 & 0.0343 & 0.9522 & 1.0002 & 0.0342 & 0.0344 & 0.9498 & 1.0065 \\
& & $\beta_2$ & -1.4996 & 0.0480 & 0.0481 & 0.9511 & -1.4996 & 0.0477 & 0.0483 & 0.9477 & 1.0069 \\
\cmidrule(lr){2-12}
& \multirow{3}{*}{$1000$} & $\beta_0$ & 0.4991 & 0.1617 & 0.1633 & 0.9495 & 0.4991 & 0.1612 & 0.1637 & 0.9477 & 1.0033 \\
& & $\beta_1$ & 1.0000 & 0.0243 & 0.0243 & 0.9496 & 1.0000 & 0.0242 & 0.0243 & 0.9481 & 1.0032 \\
& & $\beta_2$ & -1.5001 & 0.0339 & 0.0340 & 0.9503 & -1.5001 & 0.0338 & 0.0340 & 0.9488 & 1.0034 \\

\midrule
\multirow{12}{*}{\textbf{std-Exp(1)}}
& \multirow{3}{*}{$100$} & $\beta_0$ & 0.4938 & 0.5149 & 0.5180 & 0.9479 & 0.4989 & 0.3489 & 0.3591 & 0.9424 & 1.4760 \\
& & $\beta_1$ & 1.0001 & 0.0771 & 0.0779 & 0.9438 & 0.9991 & 0.0544 & 0.0560 & 0.9399 & 1.4180 \\
& & $\beta_2$ & -1.4996 & 0.1078 & 0.1113 & 0.9436 & -1.4997 & 0.0717 & 0.0751 & 0.9391 & 1.5040 \\
\cmidrule(lr){2-12}
& \multirow{3}{*}{$300$} & $\beta_0$ & 0.4980 & 0.2940 & 0.2966 & 0.9456 & 0.5007 & 0.2044 & 0.2095 & 0.9416 & 1.4390 \\
& & $\beta_1$ & 0.9997 & 0.0441 & 0.0445 & 0.9449 & 0.9989 & 0.0319 & 0.0328 & 0.9428 & 1.3860 \\
& & $\beta_2$ & -1.5006 & 0.0617 & 0.0622 & 0.9517 & -1.4997 & 0.0422 & 0.0426 & 0.9472 & 1.4650 \\
\cmidrule(lr){2-12}
& \multirow{3}{*}{$500$} & $\beta_0$ & 0.4987 & 0.2286 & 0.2291 & 0.9512 & 0.4990 & 0.1602 & 0.1622 & 0.9460 & 1.4270 \\
& & $\beta_1$ & 1.0003 & 0.0343 & 0.0343 & 0.9469 & 0.9997 & 0.0249 & 0.0250 & 0.9500 & 1.3760 \\
& & $\beta_2$ & -1.4998 & 0.0480 & 0.0481 & 0.9481 & -1.4997 & 0.0330 & 0.0333 & 0.9482 & 1.4520 \\
\cmidrule(lr){2-12}
& \multirow{3}{*}{$1000$} & $\beta_0$ & 0.5000 & 0.1615 & 0.1619 & 0.9490 & 0.4997 & 0.1146 & 0.1164 & 0.9440 & 1.4090 \\
& & $\beta_1$ & 1.0001 & 0.0243 & 0.0241 & 0.9513 & 0.9998 & 0.0178 & 0.0177 & 0.9504 & 1.3600 \\
& & $\beta_2$ & -1.4999 & 0.0339 & 0.0336 & 0.9471 & -1.4998 & 0.0236 & 0.0237 & 0.9500 & 1.4330 \\

\midrule
\multirow{12}{*}{\textbf{std-Weibull(0.6,1)}}
& \multirow{3}{*}{$100$} & $\beta_0$ & 0.5047 & 0.5016 & 0.5191 & 0.9491 & 0.5112 & 0.2728 & 0.3222 & 0.9251 & 1.8390 \\
& & $\beta_1$ & 0.9992 & 0.0752 & 0.0776 & 0.9408 & 1.0000 & 0.0438 & 0.0506 & 0.9228 & 1.7160 \\
& & $\beta_2$ & -1.5004 & 0.1051 & 0.1104 & 0.9435 & -1.5001 & 0.0553 & 0.0655 & 0.9260 & 1.9020 \\
\cmidrule(lr){2-12}
& \multirow{3}{*}{$300$} & $\beta_0$ & 0.4946 & 0.2918 & 0.2973 & 0.9456 & 0.4981 & 0.1715 & 0.1790 & 0.9438 & 1.7010 \\
& & $\beta_1$ & 1.0004 & 0.0438 & 0.0442 & 0.9496 & 1.0001 & 0.0274 & 0.0283 & 0.9412 & 1.6020 \\
& & $\beta_2$ & -1.4993 & 0.0612 & 0.0610 & 0.9517 & -1.5001 & 0.0350 & 0.0361 & 0.9479 & 1.7520 \\
\cmidrule(lr){2-12}
& \multirow{3}{*}{$500$} & $\beta_0$ & 0.4992 & 0.2262 & 0.2271 & 0.9502 & 0.4981 & 0.1372 & 0.1395 & 0.9499 & 1.6490 \\
& & $\beta_1$ & 0.9997 & 0.0340 & 0.0341 & 0.9495 & 0.9996 & 0.0218 & 0.0220 & 0.9518 & 1.5600 \\
& & $\beta_2$ & -1.4999 & 0.0475 & 0.0485 & 0.9473 & -1.4997 & 0.0280 & 0.0288 & 0.9457 & 1.6950 \\
\cmidrule(lr){2-12}
& \multirow{3}{*}{$1000$} & $\beta_0$ & 0.5019 & 0.1611 & 0.1634 & 0.9461 & 0.5010 & 0.1014 & 0.1022 & 0.9480 & 1.5880 \\
& & $\beta_1$ & 0.9997 & 0.0242 & 0.0244 & 0.9468 & 0.9998 & 0.0160 & 0.0162 & 0.9496 & 1.5090 \\
& & $\beta_2$ & -1.4997 & 0.0338 & 0.0341 & 0.9468 & -1.4996 & 0.0208 & 0.0212 & 0.9489 & 1.6280 \\

\bottomrule
\end{tabular}%
}
\end{table}

\begin{table}[htbp]
\centering
\caption{Empirical expected length ratio $R_n$ of (14) of the $\hat\beta_{OLS,j}$ and $\hat\beta_{SLS,j}$ based CIs across sample sizes.}
\resizebox{\textwidth}{!}{
\begin{tabular}{llccccccccc}
\toprule
Model, error distribution & Parameter 
& 50 & 100 & 300 & 500 & 1000 & 5000 & 10000 & 20000 & 50000 \\
\hline
\textbf{Model 1, Exp(1)} 
& $\beta_0$ & 1.456 & 1.424 & 1.401 & 1.384 & 1.369 & 1.352 & 1.354 & 1.353 & 1.351 \\
& $\beta_1$ & 1.542 & 1.505 & 1.477 & 1.457 & 1.438 & 1.417 & 1.419 & 1.418 & 1.415 \\
& $\beta_2$ & 1.542 & 1.505 & 1.477 & 1.457 & 1.438 & 1.417 & 1.419 & 1.418 & 1.415 \\
\hline

\textbf{Model 1, Weibull(0.6,1)} 
& $\beta_0$ & 1.826 & 1.742 & 1.629 & 1.573 & 1.529 & 1.453 & 1.429 & 1.421 & 1.397 \\
& $\beta_1$ & 2.088 & 1.959 & 1.796 & 1.715 & 1.653 & 1.550 & 1.517 & 1.507 & 1.475 \\
& $\beta_2$ & 2.089 & 1.958 & 1.795 & 1.715 & 1.653 & 1.550 & 1.517 & 1.507 & 1.475 \\
\hline

\textbf{Model 2, Exp(1)} 
& $\beta_0$ & 1.503 & 1.475 & 1.434 & 1.414 & 1.402 & 1.390 & 1.387 & 1.382 & 1.382 \\
& $\beta_1$ & 1.442 & 1.415 & 1.382 & 1.365 & 1.355 & 1.343 & 1.341 & 1.337 & 1.337 \\
& $\beta_2$ & 1.532 & 1.501 & 1.460 & 1.439 & 1.426 & 1.412 & 1.409 & 1.404 & 1.404 \\
\hline

\textbf{Model 2, Weibull(0.6,1)} 
& $\beta_0$ & 1.947 & 1.829 & 1.708 & 1.654 & 1.590 & 1.497 & 1.480 & 1.454 & 1.443 \\
& $\beta_1$ & 1.808 & 1.711 & 1.608 & 1.563 & 1.511 & 1.433 & 1.419 & 1.397 & 1.388 \\
& $\beta_2$ & 2.016 & 1.895 & 1.757 & 1.700 & 1.630 & 1.528 & 1.509 & 1.482 & 1.470 \\
\bottomrule
\end{tabular}}
\end{table}

\begin{table}[p]
\centering
\caption{Power function comparison for the OLSE and SLSE based tests of the hypotheses in (8) for Model 1.}
\label{tab:power-model1}
\scriptsize
\setlength{\tabcolsep}{3.2pt}
\renewcommand{\arraystretch}{0.88}

\resizebox{\textwidth}{!}{%
\begin{tabular}{@{}cc|cc|cc|cc|cc@{}}
\toprule
& & \multicolumn{2}{c|}{$n=100$}
& \multicolumn{2}{c|}{$n=300$}
& \multicolumn{2}{c|}{$n=500$}
& \multicolumn{2}{c}{$n=1000$} \\
\cmidrule(lr){3-4} \cmidrule(lr){5-6} \cmidrule(lr){7-8} \cmidrule(lr){9-10}
Parameter & Test value $\beta_{j0}$
& OLS & SLS & OLS & SLS & OLS & SLS & OLS & SLS \\
\midrule
{$\displaystyle\varepsilon \sim \mbox{std-Exp}(1)$}\\
\midrule
\multirow{3}{*}{$\beta_0$}
& 0.50 & 0.0538 & 0.0628 & 0.0488 & 0.0472 & 0.0483 & 0.0509 & 0.0554 & 0.0530 \\
& 0.72 & 0.1194 & 0.2000 & 0.2385 & 0.3981 & 0.3579 & 0.5852 & 0.5884 & 0.8366 \\
& 0.94 & 0.3104 & 0.5162 & 0.6665 & 0.8975 & 0.8668 & 0.9827 & 0.9880 & 0.9999 \\
\midrule

\multirow{3}{*}{$\beta_1$}
& 1.00 & 0.0544 & 0.0564 & 0.0515 & 0.0453 & 0.0499 & 0.0499 & 0.0528 & 0.0507 \\
& 1.02 & 0.2238 & 0.4272 & 0.5160 & 0.8396 & 0.7381 & 0.9637 & 0.9576 & 0.9993 \\
& 1.04 & 0.6434 & 0.9264 & 0.9752 & 0.9998 & 0.9990 & 1.0000 & 1.0000 & 1.0000 \\
\midrule

\multirow{3}{*}{$\beta_2$}
& -1.50 & 0.0530 & 0.0690 & 0.0521 & 0.0518 & 0.0526 & 0.0505 & 0.0546 & 0.0510 \\
& -1.45 & 0.1704 & 0.3376 & 0.4165 & 0.7239 & 0.6011 & 0.8971 & 0.8817 & 0.9950 \\
& -1.40 & 0.5188 & 0.8328 & 0.9270 & 0.9978 & 0.9924 & 0.9999 & 1.0000 & 1.0000 \\
\midrule
$\displaystyle\varepsilon \sim \mbox{std-Weibull}(0.6,1)$\\
\midrule
\multirow{3}{*}{$\beta_0$}
& 0.50 & 0.0576 & 0.0604 & 0.0544 & 0.0500 & 0.0530 & 0.0500 & 0.0492 & 0.0512 \\
& 0.72 & 0.1544 & 0.3066 & 0.2900 & 0.5040 & 0.4040 & 0.6920 & 0.5940 & 0.8940 \\
& 0.94 & 0.3636 & 0.6712 & 0.6680 & 0.9440 & 0.8800 & 0.9840 & 0.9900 & 1.0000 \\
\midrule

\multirow{3}{*}{$\beta_1$}
& 1.00 & 0.0480 & 0.0514 & 0.0510 & 0.0506 & 0.0570 & 0.0496 & 0.0484 & 0.0504 \\
& 1.02 & 0.2580 & 0.6516 & 0.5400 & 0.9350 & 0.7610 & 0.9890 & 0.9550 & 0.9990 \\
& 1.04 & 0.6862 & 0.9820 & 0.9740 & 0.9980 & 0.9960 & 1.0000 & 1.0000 & 1.0000 \\
\midrule

\multirow{3}{*}{$\beta_2$}
& -1.50 & 0.0504 & 0.0588 & 0.0452 & 0.0486 & 0.0496 & 0.0498 & 0.0540 & 0.0554 \\
& -1.45 & 0.2150 & 0.5516 & 0.4600 & 0.8920 & 0.6000 & 0.9520 & 0.9020 & 1.0000 \\
& -1.40 & 0.5660 & 0.9510 & 0.9260 & 0.9980 & 0.9920 & 1.0000 & 1.0000 & 1.0000 \\

\bottomrule
\end{tabular}%
}

\end{table}

\begin{table}[p]
\centering
\caption{Power function comparison for the OLSE and SLSE based tests of the hypotheses in (8) for Model 2.}
\label{tab:power-model2}
\scriptsize
\setlength{\tabcolsep}{3.2pt}
\renewcommand{\arraystretch}{0.88}

\resizebox{\textwidth}{!}{%
\begin{tabular}{@{}cc|cc|cc|cc|cc@{}}
\toprule
& & \multicolumn{2}{c|}{$n=100$}
& \multicolumn{2}{c|}{$n=300$}
& \multicolumn{2}{c|}{$n=500$}
& \multicolumn{2}{c}{$n=1000$} \\
\cmidrule(lr){3-4} \cmidrule(lr){5-6} \cmidrule(lr){7-8} \cmidrule(lr){9-10}
Parameter & Test value $\beta_{j0}$
& OLS & SLS & OLS & SLS & OLS & SLS & OLS & SLS \\
\midrule
$\displaystyle\varepsilon \sim \mbox{std-Exp}(1)$\\
\midrule

\multirow{3}{*}{$\beta_0$}
& 0.50 & 0.0580 & 0.0640 & 0.0478 & 0.0529 & 0.0508 & 0.0549 & 0.0490 & 0.0512 \\
& 0.72 & 0.0854 & 0.1182 & 0.1239 & 0.2078 & 0.1651 & 0.2960 & 0.2873 & 0.5025 \\
& 0.94 & 0.1586 & 0.2742 & 0.3407 & 0.5911 & 0.4944 & 0.7880 & 0.7823 & 0.9662 \\
\midrule

\multirow{3}{*}{$\beta_1$}
& 1.00 & 0.0504 & 0.0612 & 0.0512 & 0.0564 & 0.0497 & 0.0554 & 0.0494 & 0.0523 \\
& 1.02 & 0.0642 & 0.0846 & 0.0801 & 0.1124 & 0.0916 & 0.1399 & 0.1345 & 0.2167 \\
& 1.04 & 0.0978 & 0.1446 & 0.1586 & 0.2614 & 0.2234 & 0.3771 & 0.3832 & 0.6170 \\
\midrule

\multirow{3}{*}{$\beta_2$}
& -1.50 & 0.0508 & 0.0564 & 0.0531 & 0.0543 & 0.0485 & 0.0546 & 0.0538 & 0.0483 \\
& -1.45 & 0.0768 & 0.1148 & 0.1251 & 0.2196 & 0.1767 & 0.3226 & 0.3155 & 0.5648 \\
& -1.40 & 0.1572 & 0.2890 & 0.3637 & 0.6592 & 0.5437 & 0.8559 & 0.8429 & 0.9893 \\
\midrule
$\displaystyle\varepsilon \sim \mbox{std-Weibull}(0.6,1)$\\
\midrule

\multirow{3}{*}{$\beta_0$}
& 0.50 & 0.0562 & 0.0728 & 0.0580 & 0.0520 & 0.0480 & 0.0660 & 0.0520 & 0.0460 \\
& 0.72 & 0.0962 & 0.1850 & 0.1640 & 0.3260 & 0.1880 & 0.4160 & 0.3200 & 0.5880 \\
& 0.94 & 0.1860 & 0.4370 & 0.4140 & 0.7560 & 0.5300 & 0.8600 & 0.7920 & 0.9920 \\
\midrule

\multirow{3}{*}{$\beta_1$}
& 1.00 & 0.0582 & 0.0786 & 0.0680 & 0.0680 & 0.0540 & 0.0520 & 0.0580 & 0.0400 \\
& 1.02 & 0.0822 & 0.1272 & 0.0820 & 0.1470 & 0.1080 & 0.1720 & 0.1530 & 0.2690 \\
& 1.04 & 0.1238 & 0.2198 & 0.1940 & 0.3780 & 0.2400 & 0.4720 & 0.3640 & 0.7020 \\
\midrule

\multirow{3}{*}{$\beta_2$}
& -1.50 & 0.0508 & 0.0684 & 0.0480 & 0.0540 & 0.0640 & 0.0620 & 0.0460 & 0.0540 \\
& -1.45 & 0.0760 & 0.1616 & 0.1300 & 0.3080 & 0.1660 & 0.4520 & 0.3480 & 0.6940 \\
& -1.40 & 0.1712 & 0.4680 & 0.3960 & 0.8120 & 0.5340 & 0.9340 & 0.8580 & 0.9920 \\

\bottomrule
\end{tabular}%
}
\end{table}

\clearpage


\section{Real data application}

In this section, we apply the SLS method to two real data examples that have been used in the literature.

\subsection{Australian Institute of Sports (AIS) – Biomedicine}

The Australian Institute of Sports (AIS) dataset contains biomedical measurements of 706 athletes from 12 sports, collected in 1991 \citep{telford1991sex,cook1994regression}. A subset of 202 observations, available in the R sn package, has been widely used for modeling with skewed error distributions. Following the approach of Nandy et al. (2025), we model plasma ferritin concentration (Fe) using body mass index (BMI) and lean body mass (LBM) as covariates:
\[
\text{Fe}_i = \beta_0 + \beta_1 \cdot \text{BMI}_i + \beta_2 \cdot \text{LBM}_i + \varepsilon_i,\quad i=1,2,3,...,202.
\]

We first fitted the model using the OLS method and obtained the residuals. The histogram and Q-Q plot of the residuals in Figure 1 clearly indicate the right-skewness of the error distribution. Subsequently, we fitted the model again using the SLS method.
\begin{figure}[htbp]
  \centering
  \includegraphics[width=0.8\textwidth]{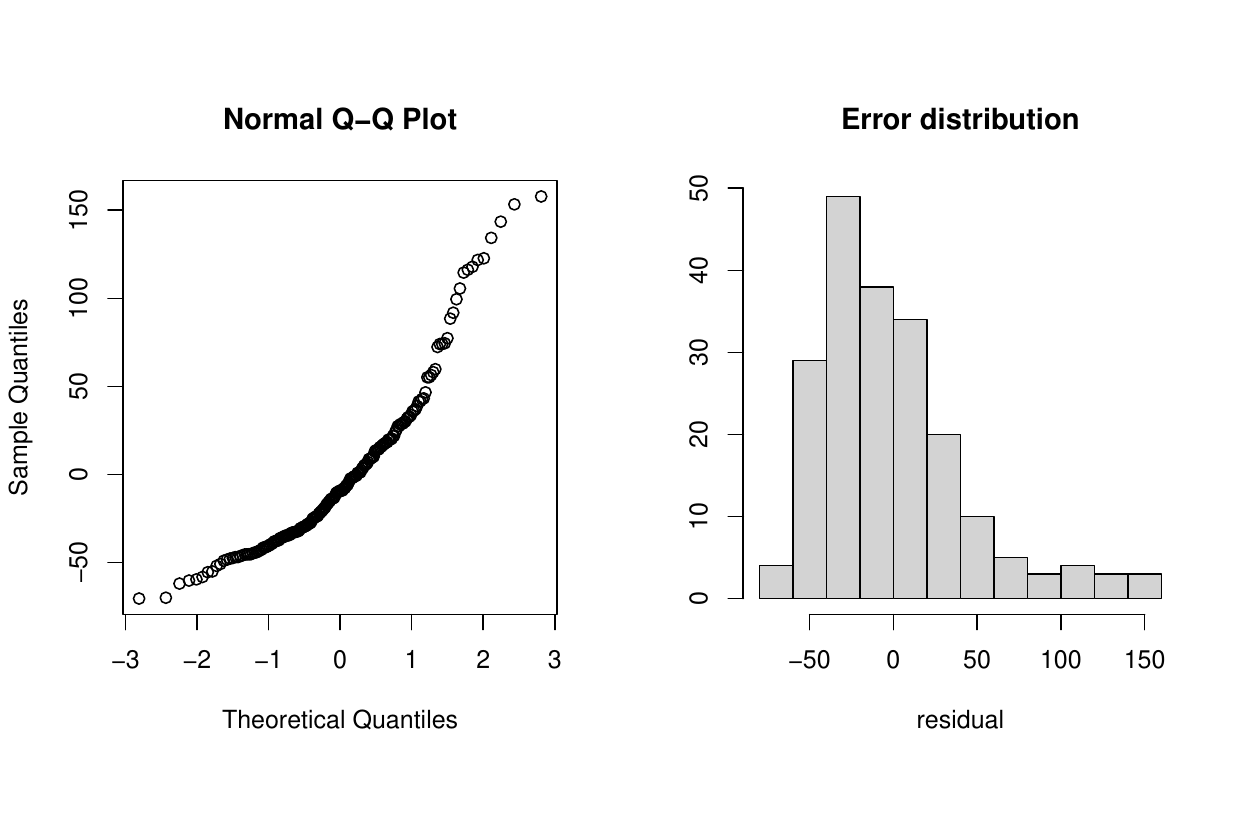}
  \caption{The OLS residual distribution of the AIS dataset.}
  \label{fig:ais_error}
\end{figure}

Tables 7 and 8 present the estimates, standard errors, and 95\% asymptotic confidence intervals for the intercept $\beta_0$ and coefficients $\beta_1$ and $\beta_2$ using the OLSEs and SLSEs. In all cases, the SLSE produces shorter CIs, indicating improved estimation precision. For instance, the interval length for the intercept decreases from 101.34 (OLSE) to 74.35 (SLSE), for $\beta_1$ (BMI) from 6.20 to 4.52, and for $\beta_2$ (LBM) from 1.36 to 0.99. These reductions reflect the variance-reducing nature of the SLSE. 
Although the SLSE point estimates differ from their OLSE counterparts, all estimates are within the corresponding CIs. However, a striking point is that the OLSE of the coefficient $\beta_1$ of BMI is insignificant at $5\%$ level because the corresponding $95\%$ CI covers zero, while the coefficient is significant by the SLSE. This means that based on the OLSE one would conclude that BMI has no impact on plasma ferritin concentration, while it does have impact based on the SLSE.

\begin{table}[htbp]
\centering
\caption{The OLSE estimates, standard errors, 95\% asymptotic CIs, and CI lengths for the ASI dataset.}
\medskip
\small
\begin{tabular}{@{}lcccc@{}}
\toprule
Parameter & Estimate & SE & 95\% CI & CI Length \\
\midrule
$\beta_0$ (Intercept) & -30.8571 & 25.8511 & [-81.5253, 19.8110] & 101.3363 \\
$\beta_1$ (BMI)       &   2.5653 &  1.5816 & [-0.5347, 5.6652] &   6.1999 \\
$\beta_2$ (LBM)       &   0.7529 &  0.3466 & [0.0737, 1.4322] &   1.3585 \\
\bottomrule
\end{tabular}
\end{table}

\begin{table}[htbp!]
\centering
\caption{The SLSE estimates, standard errors, 95\% asymptotic CIs, and CI lengths for the ASI dataset.}
\medskip
\small
\begin{tabular}{@{}lcccc@{}}
\toprule
Parameter & Estimate & SE & 95\% CI & CI Length \\
\midrule
$\beta_0$ (Intercept) & -12.6490 & 18.9672 & [-49.8247, 24.5267] & 74.3514 \\
$\beta_1$ (BMI)       &   2.3743 &  1.1530 & [0.1144, 4.6342] &  4.5198 \\
$\beta_2$ (LBM)       &   0.5398 &  0.2526 & [0.0447, 1.0350] &  0.9904 \\
\bottomrule
\end{tabular}
\end{table}
\FloatBarrier

\subsection{Health insurance}

A publicly available dataset containing information on 1,338 individuals from the USA was used to compare the performance of the OLSE and SLSE for statistical inference \citep{choi2018medicalcost}. The dataset includes variables such as age, sex, body mass index (BMI), number of children, smoking status, region of residence, and medical charges. In this analysis, medical charges were treated as the response variable. All predictor variables were standardized to improve computational efficiency. The main goal was to estimate the regression coefficients ($\beta$) and their variances using both OLS and SLS, and to compare the resulting confidence intervals.

\begin{table}[htbp]
\centering
\caption{Description of the variables in the health insurance dataset.}
\smallskip
\begin{tabular}{@{}llll@{}}
\toprule
\textbf{Variable} & \textbf{Description} & \textbf{Type} & \textbf{Values} \\
\midrule
\texttt{age} & Age of primary beneficiary & quantitative & -- \\
\texttt{sex} & Gender of insurance contractor & qualitative & \texttt{female}, \texttt{male} \\
\texttt{bmi} & Body Mass Index (BMI) & quantitative & -- \\
\texttt{children} & Number of children covered by insurance & qualitative & \texttt{0, 1, 2, ...} \\
\texttt{smoker} & Smoking status & qualitative & \texttt{yes}, \texttt{no} \\
\texttt{region} & Residential area in the U.S. & qualitative & \texttt{north/south-east/west} \\
\texttt{charges} & Individual medical costs billed by insurance & quantitative & -- \\
\bottomrule
\end{tabular}
\end{table}

Since the dataset includes categorical variables, appropriate encoding was applied before fitting the models. Specifically, sex was encoded as 0 for female and 1 for male; smoker was coded as 0 for yes and 1 for no; and region was numerically encoded as 0 for northwest, 1 for northeast, 2 for southwest, and 3 for southeast.

Similarly to the previous example, a linear model was first fitted using the OLS method. Again, the OLS residual plots in Figure 2 clearly indicate the right-skewness of the error distribution. Hence, the model was fitted again using the SLS method.
\vskip -0.42 cm
\begin{figure}[htbp]
  \centering  \includegraphics[width=0.75\textwidth]{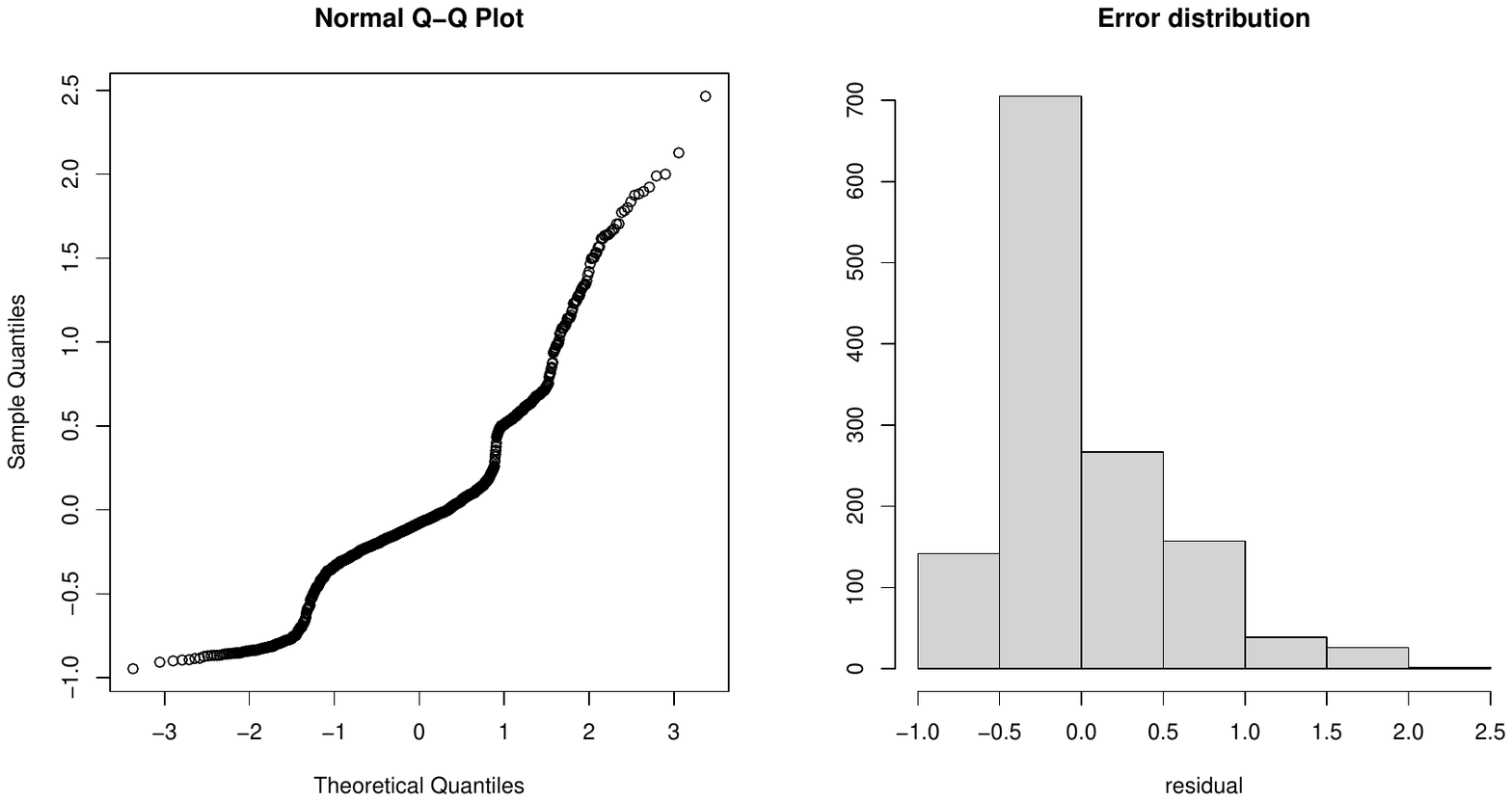}
  \vskip -1 cm
  \caption{The OLS residual distribution for the health insurance data.}
  \label{fig:medi_error}
\end{figure}

Tables 10 and 11 show the OLSE and SLSE based estimates of the  regression coefficients, standard errors, 95\% asymptotic confidence intervals, and interval lengths across the six covariates. In all cases, the SLSE produced narrower CIs than the OLSE, indicating greater precision. For instance, from Table 10, the CI for the age coefficient under the OLSE is [0.2711, 0.3252] with a length of 0.0541, while, as shown in Table 11, the SLSE yields a tighter interval of [0.2878, 0.3190], with a length of just 0.0312.

This trend is consistent across other predictors, including sex, BMI, number of children, smoker status, and region. Notably, the estimated effect of BMI (correlation of 0.2 with  charges) under the SLSE is substantially smaller (0.0722) compared to the OLSE (0.1697), yet the SLSE based CI remains narrower. This suggests that the SLSE may be less influenced by multicollinearity. A similar pattern is observed with the smoker variable (correlation of 0.79 with charges), where both methods detect a strong negative effect, but the SLSE estimate is more moderate (-0.5950 vs. -0.7956). Similarly to the previous example, the OLSE indicates that regional variation has weak or no impact on medical charges, while it has a clearly negative impact by the SLSE.

Overall, these results support the use of the SLSE for inference in this context. The tighter confidence intervals and more stable estimates suggest that the SLSE may offer improved robustness, particularly when the classical OLS assumptions, such as normality or homoscedasticity, are potentially violated.

\begin{table}[htbp]
\centering
\caption{The OLSE estimates, standard errors, 95\% asymptotic CIs, and CI lengths for the regression coefficients for the health insurance dataset.}
\bigskip
\begin{tabular}{@{}lcccc@{}}
\toprule
Covariate & Coef.\ Estimate & SE & 95\% CI & CI Length \\
\midrule
age       &  0.2981 & 0.0138 & [0.2711, 0.3252] & 0.0541 \\
sex       & -0.0053 & 0.0137 & [-0.0323, 0.0216] & 0.0539 \\
bmi       &  0.1697 & 0.0143 & [0.1417, 0.1977] & 0.0560 \\
children  &  0.0466 & 0.0137 & [0.0198, 0.0735] & 0.0537 \\
smoker    & -0.7956 & 0.0138 & [-0.8226, -0.7687] & 0.0539 \\
region    & -0.0278 & 0.0142 & [-0.0557, 0.0000] & 0.0557 \\
\bottomrule
\end{tabular}
\end{table}


\begin{table}[htbp]
\centering
\caption{The SLSE estimates, standard errors, 95\% asymptotic CIs, and CI lengths for the regression coefficients for the health insurance dataset.}
\bigskip
\small
\begin{tabular}{@{}lcccc@{}}
\toprule
Covariate & Coef.\ Estimate & SE & 95\% CI & CI Length \\
\midrule
age       & 0.3034 & 0.0080 & [0.2878, 0.3190] & 0.0312 \\
sex       & -0.0041 & 0.0079 & [-0.0196, 0.0115] & 0.0311 \\
bmi       & 0.0722 & 0.0082 & [0.0561, 0.0884] & 0.0323 \\
children  & 0.0425 & 0.0079 & [0.0270, 0.0580] & 0.0310 \\
smoker    & -0.5950 & 0.0079 & [-0.6105, -0.5794] & 0.0311 \\
region    & -0.0177 & 0.0082 & [-0.0337, -0.0016] & 0.0321 \\
\bottomrule
\end{tabular}
\end{table}

\section*{Acknowledgments}

The authors gratefully acknowledge financial support by the Natural Sciences and Engineering Research Council of Canada (NSERC) and the University of Manitoba Graduate Fellowship (UMGF).

\section*{Disclosure Statement}
No potential conflict of interest was reported by the author(s).

\bibliographystyle{chicago}

\bibliography{SLSE-Ref}

\appendix

\section{Preliminary}

Wang and Leblanc (2008) proved the consistency and asymptotic normality of the SLSE in general nonlinear models. In the case of linear models, many of their conditions and derivations are not required and, therefore, can be substantially simplified. The following lemmas are from Amemiya (1973) or Wang and Leblanc (2008). Let $Q_n(Z,\psi)$ be a measurable function of $Z$ with respect to Lebesgue measure and continuous in $\psi$, where $\psi\in\Psi$ and $\Psi\subset\IR^p$ is compact.

\begin{lem}\label{lem1}
If, as $n \to \infty$, $\sup_\Psi|Q_n(Z,\psi)-Q(\psi)|\toas 0$, where  $Q(\psi)$ is  a non-stochastic function and has a unique minimum at $\psi_0 \in \Psi$, then $\hat{\psi}_n = argmin_{\psi \in \Psi}Q_n(Z,\psi) \toas\psi_0$
\end{lem}

\begin{lem}\label{lem2}
If, as $n \to \infty$, $\sup_\Psi|Q_n(Z,\psi)-Q(\psi)|\toas 0$ and $\psi_0$ is an interior point of $\Psi$, then for any sequence of estimators $\hat{\psi}_n \xrightarrow{\text{a.s.}}\psi_0$ it holds $Q_n(Z,\hat{\psi}_n) \toas Q_n(\psi_0)$.
\end{lem}

\section{Proof of Theorem 1}

(1) First, denote $q_i(\gamma) = \rho_i(\gamma)'W_i\rho_i(\gamma)$ and 
\[
W_i = \begin{pmatrix}
w_{i,11} &w_{i,12}\\
w_{i,21} & w_{i,22}
\end{pmatrix},
\]
where $w_{i,11} = \mu_4 - \sigma_0^4  + 4\mu_3(X_i'\beta_0) + 4\sigma_0^2(X_i'\beta_0)^2$, $w_{i,12} = w_{i,21} = -\mu_3 - 2\sigma_0^2(X_i'\beta_0)$, $w_{i,22}=w_{22} = \sigma_0^2$. Then it is easy to see that the quadratic form
\[
q_i(\gamma) = w_{i,11}(Y_i -X_i'\beta)^2 + 2w_{i,12}(Y_i - X_i'\beta)(Y_i^2 - (X_i'\beta)^2 - \sigma^2)+ w_{22} (Y_i^2 - (X_i'\beta)^2 - \sigma^2)^2.
\]
can be expanded in terms of the powers of $Y_i$ and $\norm{X}_i$ and the highest orders in this expansion are $Y_i^4$, $Y_i^3\norm{X}_i$, $Y_i^2\norm{X}_i^2$ and $\norm{X}_i^4$. Further, since $\gamma$ appears only in the coefficients of these terms and the parameter space $\Gamma$ is compact, by assumption A2 it holds
$E[\max_\Gamma q_i(\gamma)] < \infty$.
It follows from the uniform law of large numbers (ULLN) that $\sup_\Gamma|Q_n(\gamma)/n -  Q(\gamma)|\toas 0$, where $Q(\gamma) = E[q_i(\gamma)]$ (Amemiya, 1985; Jennrich, 1969). 
In order to show that  $Q(\gamma)$ has a unique minimizer at the true value $\gamma_0$, note that, for any $\gamma\in\Gamma$,
\[
Q(\gamma) = Q(\gamma_0) + E[(\rho_i(\gamma)-\rho_i(\gamma_0))'W_i(\rho_i(\gamma)-\rho_i(\gamma_0))].
\]
It follows that $Q(\gamma) \geq Q(\gamma_0)$, with equality holding if and only if
\begin{equation}\label{unique}
E[(\rho_i(\gamma)-\rho_i(\gamma_0))'W_i(\rho_i(\gamma)-\rho_i(\gamma_0))]=0. 
\end{equation}
Since $W_i$ is positive definite with probability one, the above equality holds if and only if \[
\rho_i(\gamma)-\rho_i(\gamma_0) = [X_i'(\beta-\beta_0),\quad (\beta-\beta_0)'X_iX_i'(\beta+\beta_0) + \sigma^2 - \sigma_0^2)]' = 0
\]
with probability one, implying $\beta=\beta_0$ and $\sigma^2 = \sigma_0^2$.
Finally, it follows from Lemma \ref{lem1} that $\hat{\gamma}\toas \gamma_0$. 

(2) First, by Taylor expansion of the score function, we have
\begin{equation}\label{score-exp}
G_n(\hat\gamma) = G_n(\gamma_0) + H_n(\bar\gamma)(\hat\gamma - \gamma_0),
\end{equation}
where $\norm{\bar\gamma - \gamma_0}\leq \norm{\hat\gamma- \gamma_0}$, 
\begin{align*}
G_n(\gamma) &= \frac{1}{2}\delg Q_n(\gamma)\\
&= \sum_{i=1}^{n}\delg\rho_i(\gamma)' W_i \rho_i(\gamma),
\end{align*} 
and
\begin{align*}
H_n(\gamma) &= \frac{1}{2}\delgg Q_n(\gamma) \\
&= \sum_{i=1}^{n} \delg\rho_i(\gamma)' W_i \delgtr\rho_i(\gamma) + (\rho_i(\ga)'W_i\otimes I_{p+1})\delgtr\mbox{vec}(\delg\rho_i(\ga)').
\end{align*} 
Similar to the proof (1), we can show that $\sup_\gamma |H_n(\gamma)/n- H(\gamma)|\toinp 0$, where $H(\gamma) = E[\delgg Q(\gamma)]$, and hence, by the ULLN and Lemma \ref{lem2}, it holds
\begin{align*}
\frac{1}{n}H_n(\bar\gamma) &\toinp H(\gamma_0)\\
&= E[\delg\rho_i(\gamma_0)' W_i \delgtr\rho_i(\gamma_0) + (\rho_i(\ga_0)'W_i\otimes I_{p+1})\delgtr\mbox{vec}(\delg\rho_i(\ga_0)']\\
&= A,
\end{align*}
where the last equality holds because the second order derivative $\delgtr\mbox{vec}(\delg\rho_i(\ga)'$ does not depend on $Y_i$ and hence the second term has expectation zero. It follows from (\ref{score-exp}) that
\begin{equation}\label{op1}
\sqrt{n}(\hat\gamma - \gamma_0) = A^{-1}\frac{1}{\sqrt{n}}G_n(\gamma_0) + o_p(1),
\end{equation}
Further, since $E[\delg\rho_i'(\gamma_0) W_i \rho_i(\gamma_0)]= 0$, by the Central Limit Theorem, $\frac{1}{\sqrt{n}}G_n(\gamma_0)\toind N(0, B)$, where
\begin{align*}
B &= E[\delg\rho_i(\gamma_0)' W_i \rho_i(\gamma_0)\rho_i(\gamma_0)' W_i \delgtr\rho_i(\gamma_0)]\\
&= E[\delg\rho_i(\gamma_0)' W_i E[\rho_i(\gamma_0)\rho_i(\gamma_0)'|X_i] W_i \delgtr\rho_i(\gamma_0)]\\
&= E[\delg\rho_i(\gamma_0)' W_i W_i^{-1}W_i \delgtr\rho_i(\gamma_0)]\\
&= A.
\end{align*}
Thus, it follows from (\ref{op1}) that $\sqrt{n}(\hat\gamma - \gamma_0)\toind N(0, A^{-1})$.
Finally, using the expression
\[
\nabla\rho'_i(\gamma_0) = -\begin{pmatrix}
X_i & 2(X_i'\beta_0) X_i \\
    0 & 1
\end{pmatrix},
\]
we have
\[
A  = \eta
\begin{pmatrix}
(\mu_4 - \sigma_0^4 )M_{xx}   &-\mu_3 \mu_x\\
- \mu_3\mu_x'  &\sigma_0^2
\end{pmatrix}.
\]
Therefore the asymptotic covariance matrix $C=A^{-1}$ can be obtained by using the block matrix inversion formula.

\section{Proof of Theorem 2}

(1). First, since $\mu_3\neq 0$, by A1 we have $0< \xi <1$. Further, since 
\[
V(\hat\beta_{SLS})^{-1} = \frac{1}{\sigma_0^2(1-\xi)}(M_{xx} -\xi\mu_x\mu_x')
\]
and $V(\hat\beta_{OLS})^{-1} = \frac{1}{\sigma_0^2}M_{xx}$, we have
\begin{align*}
\sigma_0^2[V(\hat\beta_{SLS})^{-1} - V(\hat\beta_{OLS})^{-1}] 
&= \frac{\xi}{1-\xi}(M_{xx} -\mu_x\mu_x')\\
&= \frac{\xi}{1-\xi}\Sigma_{xx}
\end{align*}
which is nonnegative definite, where $\Sigma_{xx}$ is the covariance matrix of $X$. It follows that $V(\hat\beta_{OLS}) - V(\hat\beta_{SLS})$ is nonnegative definite.  Further, since
\[
\Sigma_{xx} = M_{xx}^{1/2}(I - M_{xx}^{-1/2}\mu_x\mu_x'M_{xx}^{-1/2})M_{xx}^{1/2}
\]
and $M_{xx}^{-1/2}\mu_x\mu_x'M_{xx}^{-1/2}$ has the same nonzero eigenvalue as 
$\mu'_xM^{-1}_{xx}\mu_x$, $\Sigma_{xx}$, and therefore $V(\hat\beta_{OLS}) - V(\hat\beta_{SLS})$, is positive definite if and only if $\mu'_xM^{-1}_{xx}\mu_x\neq 1$.

(2). First, since
\[
1 - \mu'_xM^{-1}_{xx}\mu_x = E[(1 - \mu'_xM^{-1}_{xx}X)^2] \geq 0,
\]
we have $\mu'_xM^{-1}_{xx}\mu_x\leq 1$.  It follows that
\[
V(\hat\sigma^2_{OLS}) - V(\hat\sigma^2_{SLS}) = (\mu_4-\sigma_0^4)\left(1 - \frac{1-\xi}{1-\xi \mu'_xM^{-1}_{xx}\mu_x}\right) \geq 0
\]
with equality holding if and only if $\mu'_xM^{-1}_{xx}\mu_x = 1$.

\section{Derivation of the gradient $G(\gamma)$ and Hessian matrix $H(\gamma)$ of $Q_n(\gamma)$}

We derive the gradient and Hessian for the given data $y_i, x_i, i=1,2,...,n$. 
To evaluate the gradient, we first have
\[
\delg\rho_i'(\gamma) = -\begin{pmatrix}
x_i   &2(x_i'\beta)x_i\\
0 &1
\end{pmatrix},
\]
\[
\delg\rho_i'(\gamma)W_i = -\eta
\begin{pmatrix}
[\mu_4 - \sigma^4 + 2\mu_3(x_i'\beta)]x_i   &-\mu_3 x_i\\
- 2\sigma^2(x_i'\beta)- \mu_3  &\sigma^2
\end{pmatrix}
\]
and therefore
\[
\delg\rho_i'(\gamma)W_i\rho_i(\gamma) = -\eta
\begin{pmatrix}
[(\mu_4 - \sigma^4)(y_i - x_i'\beta) -\mu_3(y_i - x_i'\beta)^2 + \mu_3\sigma^2 ]x_i\\
\sigma^2(y_i - x_i'\beta)^2  - \sigma^4 - \mu_3(y_i - x_i'\beta)
\end{pmatrix}.
\]
Further, to evaluate the Hessian, we have
\[
\delg\rho_i'(\gamma)W_i \delgtr\rho_i(\gamma)  = \eta
\begin{pmatrix}
(\mu_4 - \sigma^4 )x_ix_i'   &-\mu_3 x_i\\
- \mu_3x_i'  &\sigma^2
\end{pmatrix}.
\]
Since
\[
\delgtr\mbox{vec}(\delg\rho_i'(\ga)) = -
\begin{pmatrix}
0   &0\\
0   &0\\
2x_ix_i'  &0\\
0    & 0
\end{pmatrix},
\]
we have
\begin{align*}
& (\rho_i'(\ga)W_i\otimes I_{p+1})\delgtr\mbox{vec}(\delg\rho_i'(\ga))\\
&= -2\eta 
\begin{pmatrix}
[\sigma^2(y_i - x_i'\beta)^2 - \sigma^4 - \mu_3 (y_i - x_i'\beta)]x_ix_i'    &0\\
0   &0
\end{pmatrix}
\end{align*}
and therefore
\begin{align*}
H_i = \eta
\begin{pmatrix}
[\mu_4 + \sigma^4  - 2\sigma^2(y_i - x_i'\beta)^2 + 2\mu_3 (y_i - x_i'\beta)]x_ix_i'    &-\mu_3 x_i\\
-\mu_3 x_i'   &\sigma^2
\end{pmatrix}.
\end{align*}

It follows that
\begin{align*}
G(\gamma) &= \sum_{i=1}^{n}\delg\rho_i'(\gamma) W_i \rho_i(\gamma)\\
&= -\eta\sum_{i=1}^{n}
\begin{pmatrix}
[(\mu_4 - \sigma^4)(y_i - x_i'\beta) -\mu_3(y_i - x_i'\beta)^2 +  \mu_3\sigma^2 ]x_i\\
\sigma^2(y_i - x_i'\beta)^2 - \sigma^4 - \mu_3(y_i - x_i'\beta) 
\end{pmatrix}
\end{align*} 
and
\begin{align*}
H(\gamma) &= \sum_{i=1}^{n} \delg\rho_i'(\gamma) W_i \delgtr\rho_i(\gamma) + (\rho_i'(\ga)W_i\otimes I_{p+1})\delgtr\mbox{vec}(\delg\rho_i'(\ga))\\
&= \eta \sum_{i=1}^{n} 
\begin{pmatrix}
[\mu_4 + \sigma^4 - 2\sigma^2(y_i - x_i'\beta)^2 + 2\mu_3 (y_i - x_i'\beta)]x_ix_i'    &-\mu_3 x_i\\
-\mu_3 x_i'   &\sigma^2
\end{pmatrix}.
\end{align*}


\end{document}